\documentclass[unnumsec,webpdf,modern,large]{oup-authoring-template}

\graphicspath{{fig/}}

\theoremstyle{thmstyleone}%
\theoremstyle{thmstyletwo}%
\theoremstyle{thmstylethree}%

\begin{document}

\journaltitle{arXiv manuscript}
\DOI{DOI HERE}
\copyrightyear{2025}
\pubyear{2025}
\access{Advance Access Publication Date: Day Month Year}
\appnotes{Original Contribution}

\firstpage{1}


\title[Causal language jumps in guidelines]{Causal language jumps in clinical practice guidelines for diabetes management}

\author[1,$\ast$]{Keling Wang \ORCID{0000-0003-1769-5104}}
\author[1]{Chang Wei}
\author[1]{Jeremy A. Labrecque}

\authormark{Keling Wang et al.}

\address[1]{\orgdiv{Department of Epidemiology}, \orgname{Erasmus MC University Medical Center Rotterdam}, \orgaddress{\street{Dr. Molewaterplein 40}, \postcode{3015GD}, \state{Rotterdam}, \country{The Netherlands}}}

\corresp[$\ast$]{Corresponding author. \href{mailto:keling.wang@erasmusmc.nl}{keling.wang@erasmusmc.nl}}

\received{Date}{0}{Year}
\revised{Date}{0}{Year}
\accepted{Date}{0}{Year}


\abstract{
Clinical practice guidelines are designed to guide clinical practice and involve causal language. Sometimes guidelines make or require stronger causal claims than those in the references they rely on, a phenomenon we refer to as ``causal language jump''. We evaluated the strength of expressed causation in diabetes guidelines and the evidence they reference to assess the pattern of jumps. 
We randomly sampled 300 guideline statements from four diabetes guidelines. We rated the causation strength in the statements and the dependence on causation in recommendations supported by these statements using existing scales. Among the causal statements, the cited original studies were similarly assessed. We also assessed how well they report target trial emulation (TTE) components as a proxy for reliability. 
Of the sampled statements, 114 (38.0\%) were causal, and 76 (66.7\%) expressed strong causation. 27.2\% (31/114) of causal guideline statements demonstrated a "causal language jump", and 34.9\% (29/83) of guideline recommendations cannot be effectively supported. Of the 53 eligible studies for TTE rating, most did not report treatment assignment and causal contrast in detail.
Our findings suggest causal language jumps were common among diabetes guidelines. 
While these jumps are sometimes inevitable, they should always be supported by good causal inference practices.
}


\maketitle

\section{}
Clinical practice guidelines (abbreviated as ``guidelines'') are ``[A] set of statements that include recommendations intended to optimize patient care that are informed by a systematic review of evidence and an assessment of the benefits and harms of alternative care options''\cite{institute_of_medicine_us_committee_on_standards_for_developing_trustworthy_clinical_practice_guidelines_clinical_2011}. Guidelines have become increasingly important in clinical decision-making since evidence-based care has gained attention \cite{fox_practice_2009}, and thus recommendations from guidelines often heavily impact actual treatment decisions or care strategies. 

Guidelines often concern an action or treatment decision. Guideline recommendations that aim to inform practitioners about giving or withholding treatment are necessarily causal in nature, because they compare the consequences of two (or more) potential treatment options. Causation, however, is difficult to demonstrate, and causal evidence often requires very careful interpretation. If the causal statements in guidelines are not aligned with the evidence on which it relies, practitioners may make suboptimal decisions affecting patient or population health. 

\begin{figure}[ht]
    \centering
    \includegraphics[width=\linewidth]{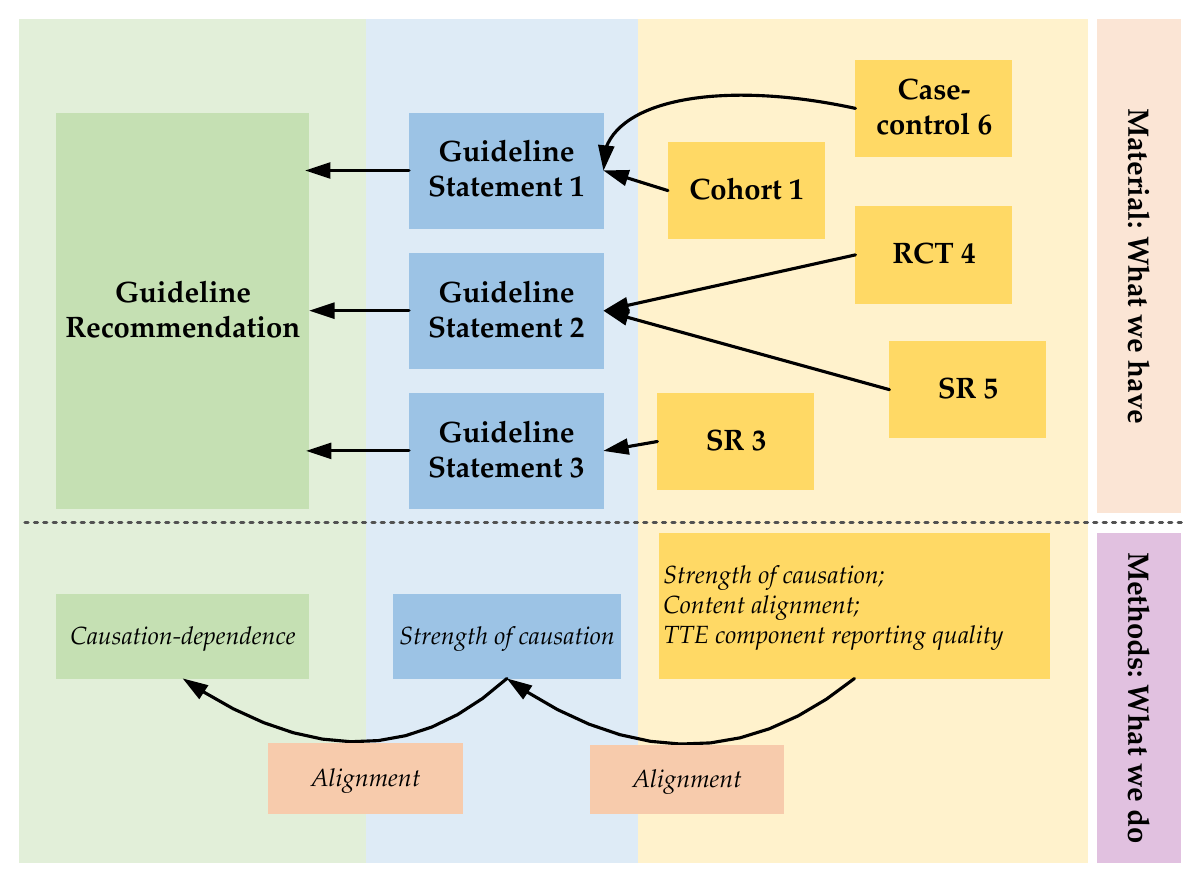}
    \smallskip
    \begin{tablenotes}
        \small SR, systematic review; RCT, randomized controlled trial; TTE, target trial emulation. \textbf{Upper panel} shows the basic structure of a guideline as well as its original studies; \textbf{Lower panel} gives an overview of our methods for the assessment of causal language jumps and misalignment along the entire causation flow.
    \end{tablenotes}
    \caption{Structure of causation flows in a clinical practice guideline and the assessment of causal language jumps along this flow}
    \label{fig:causationflow}
\end{figure}

The strength of causation expressed in guidelines should be adequately supported by the existing evidence they references. The strength of causal conclusions should flow from the original studies to the statements in guidelines that narrate evidence, and finally to the guideline recommendations that inform clinical practice (Figure~\ref{fig:causationflow}). To ensure the validity of a recommendation, all causal information should be faithfully passed in this flow.

However, the flow of the strength of causal language may not always be well-supported. The strength of causation expressed in guidelines may be stronger than that expressed in the literature. For example, observational studies often use associational language~\cite{kezios_is_2021,hernan_c-word_2018} but are still used to support causal statements in guidelines. There could also be recommendations that rely on a larger assumption set than the ones implied in their corresponding original studies. 

We use the term \textit{``causal language jump''} to refer to the situations where the causation or the dependence to causation is stronger than the previous step in the causal flow. 
In a guideline's causation flow, a causal language jump could be non-alignment of the expressed strength of causation or a gap from non-causal to causal statements when passing evidence from original studies to guideline statements and finally recommendations. Attention should be paid to this discontinuity as they can have consequences in real-world clinical decision-making. Whether and how causal language jumps are made in guidelines, however, are not known yet.

Causal language in scientific literature often consists of a set of words, modifiers and specific sentence structures to express the existence, confidence, uncertainty, magnitude and other attributes of causality~\cite{yu_detecting_2019,kaufman_there_2016,haber_causal_2022}, which is fundamental to modern epidemiology~\cite{greenland_identifiability_1986}. There have been studies assessing the use of causal language \cite{haber_causal_2022,parra_consistency_2021,grosz_taboo_2020,han_causal_2022}, finding inconsistency in expressions of causation. For example, Haber et al. \cite{haber_causal_2022} rated the causation strength expressed in abstracts and main texts of more than 1000 articles, and suggested the existence of a disconnect between the use of causal language and the implications in these articles. 
There are also published studies in the field of health communication~\cite{adams_claims_2019}, media~\cite{adams_how_2017,haber_causal_2018}, and computer science~\cite{li_nlp_2017} that note this disconnect between the intention and expression of causation. 

These studies on causal language use, however, focus on observational studies themselves while little work has been done on how causal or non-causal language can have a larger impact, for example, on guidelines. In other words, they did not assess the full ``causation flow'', but only the first stop on this flow. In order to get a full picture of causal language use and potential jumps in guidelines, we need to assess causal language not only within the guidelines or original studies themselves, but also between them. 

Therefore, we aimed to evaluate the use of causal language in guidelines and the original studies they cited, and illustrate the pattern of potential causal language jumps by checking the alignment of causal language. We focused on clinical guidelines that provided recommendations for non-pharmacological prevention, treatment and management of adults type 2 diabetes mellitus (T2DM), because the evidence for non-pharmacological treatments is well-developed for T2DM \cite{raveendran_non-pharmacological_2018} but is also more vulnerable to biases, and can potentially provide a good illustration for the issue we examined.

\section{Material \& Methods}\label{methods}

\subsection{Study design}

We followed the PRISMA-ScR~\cite{tricco_prisma_2018} guidelines with some deviations as we did not perform a standard scoping review. The protocol of this study was registered at the Open Science Framework\cite{wang_causal_2024}. 
The workflow is displayed in Figure~\ref{fig:causationflow} and is described in detail below. Briefly, (1) we selected eligible guideline recommendations and their surrounding guideline statements; (2) we randomly sampled from these guideline statements, included the recommendations they support, and assessed the strength of causal language in statements or the dependence to causal evidence in recommendations; (3) we collected the original studies these statements cited, assessed the strength of causal language and assessed the alignment between the two; and (4) we checked the reporting quality of the eligible original studies in terms of TTE components to see whether the quality sufficed to support their conclusions.

\subsection{Guideline selection}
We selected diabetes guidelines that were developed by internationally recognized academic associations and used the most. These guidelines should focus on non-pharmacological treatments or preventive strategies for adult T2DM management and related complications, and be with clear quality of evidence ratings and development methodologies. Eight chapters from four sets of guidelines~\cite{committee_3_2024,committee_5_2024,committee_8_2024,committee_10_2024,committee_13_2024,marx_2023_2023,sinclair_managing_2013,international_diabetes_federation_global_2014} were selected according to both the predefined criteria and suggestions on source of guideline sets from expert researchers in the field of diabetes mellitus. Details of these guidelines are reported in \textbf{Supplementary File S1}.

\subsection{Recommendation and guideline statement selection}

All action recommendations (Figure~\ref{fig:causationflow} leftmost column) from the guideline sections mentioned above were eligible for assessment if they focused on non-pharmacological treatment or preventive strategies of type 2 diabetes or in a general diabetic context. We manually picked all these eligible recommendations for further analysis.

These recommendations were typically surrounded and supported by a number of sentences that were used to narrate evidence and provide support to recommendations, which we referred to as \textit{guideline statements} (Figure~\ref{fig:causationflow} middle column). All the guideline statements attached to an eligible action recommendation were also considered eligible as long as they focused on non-pharmacological treatment strategies. Due to our capacity limit, we randomly drawn 300 statements and picked their associated action recommendations from all eligible ones to proceed. 

\subsection{Recommendation and guideline statement assessment}

We rated each of the 300 statements for the expressed strength of causation based on causal linking words, modifiers, and sentence tone about uncertainty. For the associated recommendation, we rated causation-dependence, \textit{i.e.}, to what extent the recommendation relied on causal evidence. This was judged upon routine clinical practice and subject-matter knowledge. Subsequently, we checked whether one recommendation can be effectively supported in terms of causation strength, by comparing its causation-dependence and all the causation strength ratings of its surrounding guideline statements.

A four-level scale, namely \textit{The Causal Implication Strength Rating Scale}  \cite{haber_causal_2022}, was used for causation and causation-dependence ratings. We rated each statement as ineligible (\textit{i}) if they were related to pharmacological treatments, no relationship expressed between variables (-1), correlational but not causal (0), weak causal (1), moderate causal (2), and strong/explicit causal (3); each recommendation as no causal evidence needed to make recommendations (0), little causal evidence needed (1), some or possible causal evidence needed (2), and definitely explicit causal evidence needed (3). Detailed explanation and examples of these ratings can be found in \textbf{File S2}.

\subsection{Original study selection and assessment}

For each guideline statement that was sampled and rated as a causal one, we extracted all eligible studies cited by this statement ((Figure~\ref{fig:causationflow} rightmost column). Eligible cited studies were intervention studies, observational studies, systematic reviews of intervention and/or observational studies, or aggregated evidence. If there was no study cited, we checked surrounding sentences to determine whether they contained a reference that might support the statement. 

For each original study, we extracted one conclusive sentence from the main text that best supported the associated guideline statement. These sentences were rated for their strength of causation using the same scale described above. Additionally, a three-level item was used to rate for the alignment of the contents of these potentially causal sentences to those in the guideline statements (``not aligned'', ``partially aligned'', and ``completely aligned''). This alignment score was used to assess whether original studies could effectively support guideline statements in terms of contents. Examples of content alignment ratings can be found in \textbf{File S2}.

After assessing causation strength and alignment in the main text, we randomly picked one original study for each guideline statement (if there existed more than one) to further assess the reporting quality. This was done in two additional steps. First, we extracted one conclusive sentence and one action recommendation sentence in the abstract. These abstract sentences were rated similarly as those in the main texts, using the same scale. By this we aimed to verify whether the conclusive sentences in abstracts carried causation strength of similar patterns and alignment compared to those from main texts.

Second, in order to assess how well the design and study details were reported to support a causal claim in the original studies as a proxy for the reliability and quality of the claim, we extracted the study components or corresponding emulated trial components for primary reports (single trials and observational studies) among these selected studies. We evaluated them in terms of assessing the reporting of (target) trial components under a Target Trial Emulation (TTE) Framework \cite{hernan_using_2016,hernan_target_2022}. This included checking the reporting of eligibility criteria, treatment strategies, assignment procedures, follow-up period, outcome, causal contrast, and analysis plan; each of them was rated as 0 (not or inappropriately reported), 1 (partially or problematically reported), or 2 (fully reported)~\cite{smit_causal_2023}. Explanations of the TTE component reporting quality scale can be found in \textbf{File S2}.

\subsection{Divergence resolving}

Ratings were done by two coauthors (KW and JAL for guideline ratings; KW and CW for original study ratings) independently. The ratings were then compared and assessed for agreement. All the discrepancies in guideline statement ratings and guideline recommendation ratings were discussed and resolved by KW, CW, and JAL together. All other discrepancies in original study sentences and reporting quality ratings were discussed and resolved by two coauthors together. Data extraction table for study characteristics and other data selection and extraction components were done by one author (KW) and checked by second (CW).

\subsection{Data synthesis and statistical considerations}\label{sec:data_synthesis}
A narrative and descriptive summary is presented. For categorical variables, the count and proportion are reported. For continuous variables, means and standard deviations (SD),  or medians and inter-quartile range (IQR) are reported accordingly. All statistical analyses were implemented using \texttt{R 4.4} in RStudio. A seed number of 114514 was used for all procedures involving random sampling.

\section{Results}

\subsection{Causal language use and alignment within guidelines}

We extracted 1175 guideline statements and 183 recommendations. Of all guideline statements, 715 (60.9\%) were followed by at least one reference. Other characteristics about word frequencies are depicted in \textbf{File S1}. 

Among the 300 sampled statements, 114 (38.0\%) were causal, of which 15 (13.2\%) were weak, 23 (20.2\%) were moderate, and 76 (66.7\%) were strong causal statements. Table~\ref{tab:causation-guideline} provides details for all the ratings.

\begin{table}[ht]
            \centering
            \caption{Strength of causation ratings for guideline statements}
            \label{tab:causation-guideline}
            \smallskip
            \begin{tabular}{lrrrr}
            \toprule
            Level of causation & Rating & No. & \% in category & \% in total \\
            \midrule
            Causal & ~ & 114 & 100\% & 38.0\% \\
            ~~~ Weak & 1 & 15 & 13.2\% & 5.00\% \\
            ~~~ Moderate & 2 & 23 & 20.2\% & 7.67\% \\
            ~~~ Strong & 3 & 76 & 66.7\% & 25.3\% \\
            \midrule
            Non-causal & ~ & 170 & 100\% & 56.7\% \\
            ~~~ Correlational & 0 & 37 & 21.8\% & 12.3\% \\
            ~~~ No relationship & -1 & 133 & 78.2\% & 44.3\% \\
            \midrule
            Ineligible & N/A & 16 & N/A & 5.33\% \\
            \midrule
            Total & ~ & 300 & ~ & 100\% \\
            \botrule 
            \end{tabular}
            \begin{tablenotes}
                \small N/A, not applicable. \% in category takes the total number of sentences in this category (causal or non-causal) as denominator.
            \end{tablenotes}
\end{table}

Following statement sampling, we extracted 83 guideline recommendations that were supported by these guideline statements. About one third (29, 34.9\%) of the recommendations could not be effectively supported by any of the surrounding statements in terms of expressed strength of causation (Table~\ref{tab:causationdep-rec-guideline}). Examples of this type of causal language jumps are provided in Table~\ref{tab:eg_causation_jump}.

\begin{table*}[ht]
    \centering
    \caption{Causation-dependence of action recommendations and alignment}\label{tab:causationdep-rec-guideline}
    \smallskip
    \begin{tabular*}{\textwidth}{@{\extracolsep{\fill}}lrrrr@{\extracolsep{\fill}}}
    \toprule
         Causation dependence & Rating & No. & No. eff. supported & No. ineff. supported\\
    \midrule
    Guideline rec. & ~ & 83 (100\%) & 54 (65.1\%) & 29 (34.9\%) \\
         ~~~~Strong & 3 & 24 (28.9\%) & 18 (21.7\%) & 6 (7.23\%) \\
         ~~~~Moderate & 2 & 23 (27.7\%) & 13 (15.7\%) & 10 (12.0\%) \\
         ~~~~Weak & 1 & 19 (22.9\%) & 6 (7.23\%) & 13 (15.7\%)\\
         ~~~~None & 0 & 17 (20.5\%) & 17 (20.5\%) & 0 (0.00\%) \\
    \midrule
    OS Abstract rec.  & ~ & 38 (100\%) & 35 (92.1\%) & 3 (7.89\%) \\
         ~~~~Strong & 3 & 8 (21.1\%) & 7 (18.4\%) & 1 (2.63\%) \\
         ~~~~Moderate & 2 & 16 (42.1\%) & 14 (36.8\%) & 2 (5.26\%) \\
         ~~~~Weak & 1 & 4 (10.5\%) & 4 (10.5\%) & 0 (0.00\%)\\
         ~~~~None & 0 & 10 (26.3\%) & 10 (26.3\%) & 0 (0.00\%)\\
    \midrule
    OS Main text rec.  & ~ & 72 (100\%) & 65 (90.3\%) & 7 (9.72\%) \\
         ~~~~Strong & 3 & 17 (23.6\%) & 14 (19.4\%) & 3 (4.17\%) \\
         ~~~~Moderate & 2 & 32 (44.4\%) & 29 (40.3\%) & 3 (4.17\%) \\
         ~~~~Weak & 1 & 18 (25.0\%) & 17 (23.6\%) & 1 (1.39\%)\\
         ~~~~None & 0 & 5 (6.94\%) & 5 (6.94\%) & 0 (0.00\%)\\
    \botrule
    \end{tabular*}
    \begin{tablenotes}
        \small Note: rec., recommendation; eff., effectively; ineff., ineffectively. Total number of recommendations excludes some studies where no recommendations nor conclusive sentences existed in abstracts or main texts. Percentages take the total number of recommendations in each corpus as denominator.
    \end{tablenotes}
\end{table*}

\begin{table*}[!ht]
\small
\centering
\caption{Different types of causal language jump and non-alignment within and beyond guidelines}\label{tab:eg_causation_jump}
\smallskip
    \begin{tabular*}{\textwidth}{@{\extracolsep{\fill}}p{0.05\textwidth}p{0.35\textwidth}p{0.1\textwidth}p{0.35\textwidth}p{0.1\textwidth}@{\extracolsep{\fill}}}
    \toprule
    Example & Text & Rating & Text & Rating\\
    \midrule
    ~ & \multicolumn{4}{l@{}@{}@{}}{\textit{1. Causal language jump between guideline recommendations and guideline statements}}  \\
    \cline{2-3}\cline{4-5}
    eg.1.1 & \textit{Guideline recommendation:} Principles of motivational interviewing should be considered to induce behavioural changes. & Moderate dependence & \textit{Guideline statement with highest causation strength rating:} Perceived susceptibility to illness and the anticipated severity of the consequences are also prominent components of patients’ motivation. & No relationship \\
    
    eg.1.2 & \textit{Guideline recommendation:} Provide an increased level of support for people with diabetes and serious mental illness through enhanced monitoring of and assistance with diabetes self-management behaviors. & Moderate dependence & \textit{Guideline statement with highest causation strength rating:} Disordered thinking and judgment can be expected to make it difficult to engage in behavior that reduces risk factors for type 2 diabetes, such as restrained eating for weight management. & Weak causation \\
    \midrule
    
    ~ & \multicolumn{4}{l@{}@{}@{}}{\textit{2. Causal language jump between guideline statements and original studies (not accounting for content alignment)}}  \\
    \cline{2-3}\cline{4-5}
    eg.2.1 & \textit{Guideline statement: }It is known that smokeless tobacco products, such as dip and chew, pose an increased risk for CVD (348). & Strong causation &  \textit{Original study sentence with highest rating:} Promoting ST product use as a way for smokers to reduce risk for smoking-related diseases is not appropriate & Weak causation, partially aligned in contents \\
    eg.2.2 & \textit{Guideline statement: } A cluster randomized trial found statistically significant increases in quit rates and long-term abstinence rates (>6 months) when smoking cessation interventions were offered through diabetes education clinics, regardless of motivation to quit at baseline (360). & Strong causation & \textit{Original study sentence with highest rating:} Implementation of OMSC in diabetes education programs was associated with a near quadrupling of the likelihood that smokers with type 2 diabetes or prediabetes would achieve long-term abstinence. & Moderate causation, fully aligned in contents\\
    \midrule
    
    ~ & \multicolumn{4}{l@{}@{}@{}}{\textit{3. Causal language jump between guideline statements and original studies (accounting for content alignment)}}  \\
    \cline{2-3}\cline{4-5}
    eg.3.1 & \textit{Guideline statement: }While post-cessation weight gain is an identified issue, studies have found that an average weight gain of 3–5 kg does not necessarily persist long term or diminish the substantial cardiovascular benefit realized from smoking cessation (337). & Strong causation & \textit{Original study sentence with highest rating:} both active and passive smoking are associated with increased risk of incident type 2 diabetes & Moderate causation, not aligned at all in contents \\
    eg.3.2 & \textit{Guideline statement: }People with CVD and T2DM are encouraged to reduce sodium intake, as this may reduce systolic BP by, on average, 5.8 mmHg in hypertensive patients and 1.9 mmHg in normotensive patients (94,95). & Moderate causation & \textit{Original study sentence with highest rating:} decreasing sodium intake reduces blood pressure in those with diabetes & Strong causation, partially aligned in contents \\    
    \botrule
         
    \end{tabular*}
    
\end{table*}

\subsection{Original study characteristics}

There were 191 original studies cited to support causal statements. Two studies were excluded from the analysis because we could not retrieve the full text. Study information and characteristics can be found at \textbf{File S3}. Of these studies, most were from the USA (109, 57.1\%), published in Diabetes Care (29, 15.2\%) or NEJM (14, 7.33\%), and published between 2017 and 2022 (93, 48.7\%). The articles received a median citation number of 232.0 (IQR 51.0 to 659.0), and 22.20 citations per year (IQR 8.75 to 68.10). Original studies were mainly RCTs (52, 27.2\%), followed by systematic reviews (with meta-analysis, 40, 20.9\%) that mainly consisted of RCTs as well. Twenty (10.5\%) studies were secondary analyses or follow-up studies of an RCT, and only fifteen (7.85\%) were observational. All the 191 studies were included in the assessment of original study main texts; after sampling, 97 studies corresponding to 97 guideline statements were further analyzed for abstracts, action recommendations, and TTE components.

\subsection{Alignment between original studies and guidelines}

Of the 114 causal statements, 30 (26.3\%) were not directly followed by any reference, and 12 (10.5\%) failed to be supported by any original studies even in the surrounding statements. The causal statements had a median of 1 original study (IQR 0 to 2, min. 0, max. 13, mean 1.64).

\begin{figure*}
    \centering
    \includegraphics[width=0.8\textwidth]{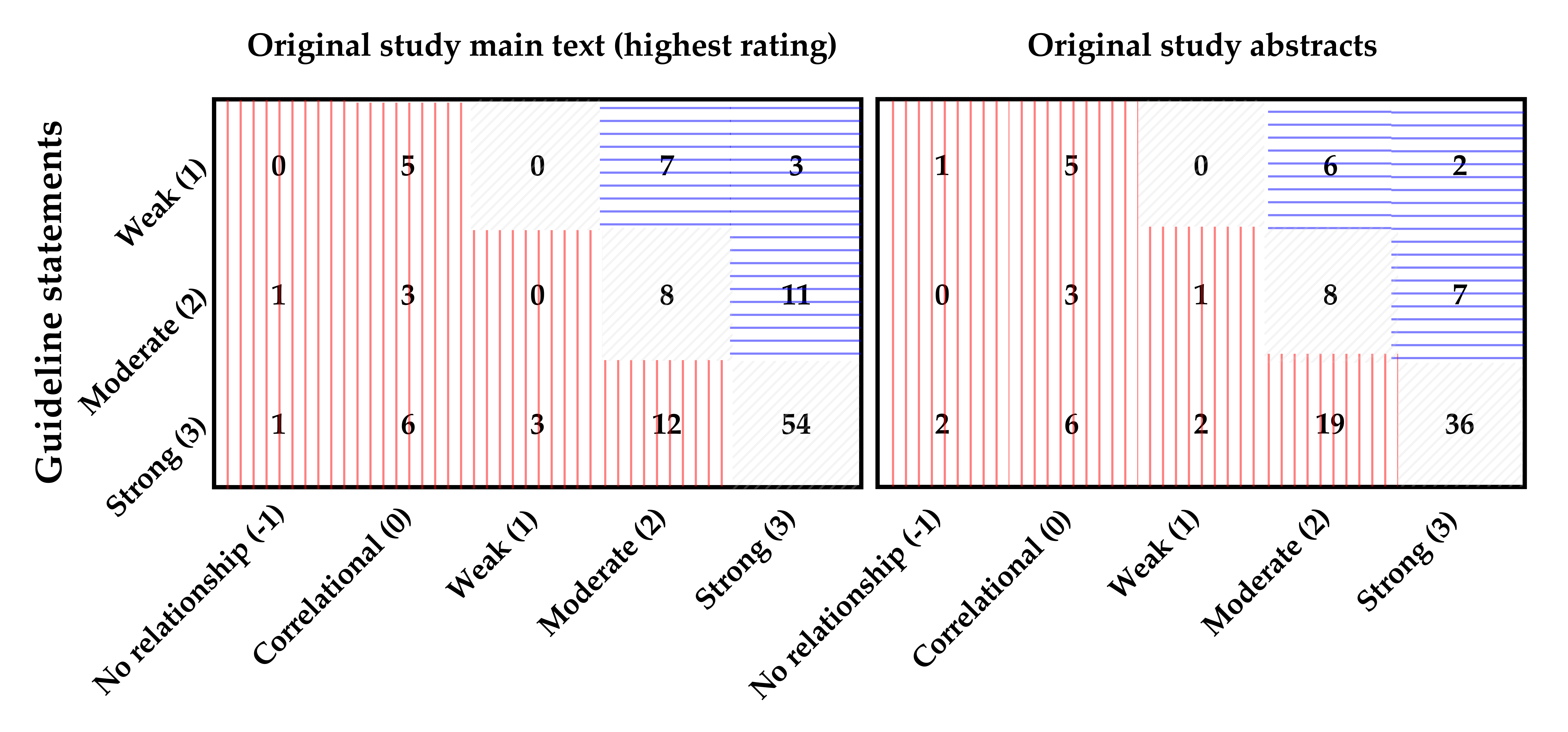}
    \begin{tablenotes}
        \small \textbf{Left panel}: Alignment of causation ratings between causal guideline statements and main-text conclusive sentences \textit{with the highest ratings} from their associated original studies; $N = 114$; \textbf{Right panel}: Alignment of causation ratings between causal guideline statements and abstract conclusive sentences from their associated original studies; $N = 98$ because some original studies do not have an eligible abstract; \\
        Numbers in cells refer to the number of items (sentences). Red cells indicate non-alignment / causal language jumps. Cells filled with red vertical line indicate items with lower ratings in original studies but higher ones in guideline statements; blue horizontal line indicate items with higher ratings in original studies and lower ones in guideline statements; gray diagonal line indicate items with same rating in both guideline statements and original studies.
    \end{tablenotes}
    \caption{Alignment of causation strength ratings in both abstracts and main texts}
    \label{fig:heatmap_ratings}
\end{figure*}

\begin{figure*}[ht]
    \centering
    \includegraphics[width=1.05\textwidth]{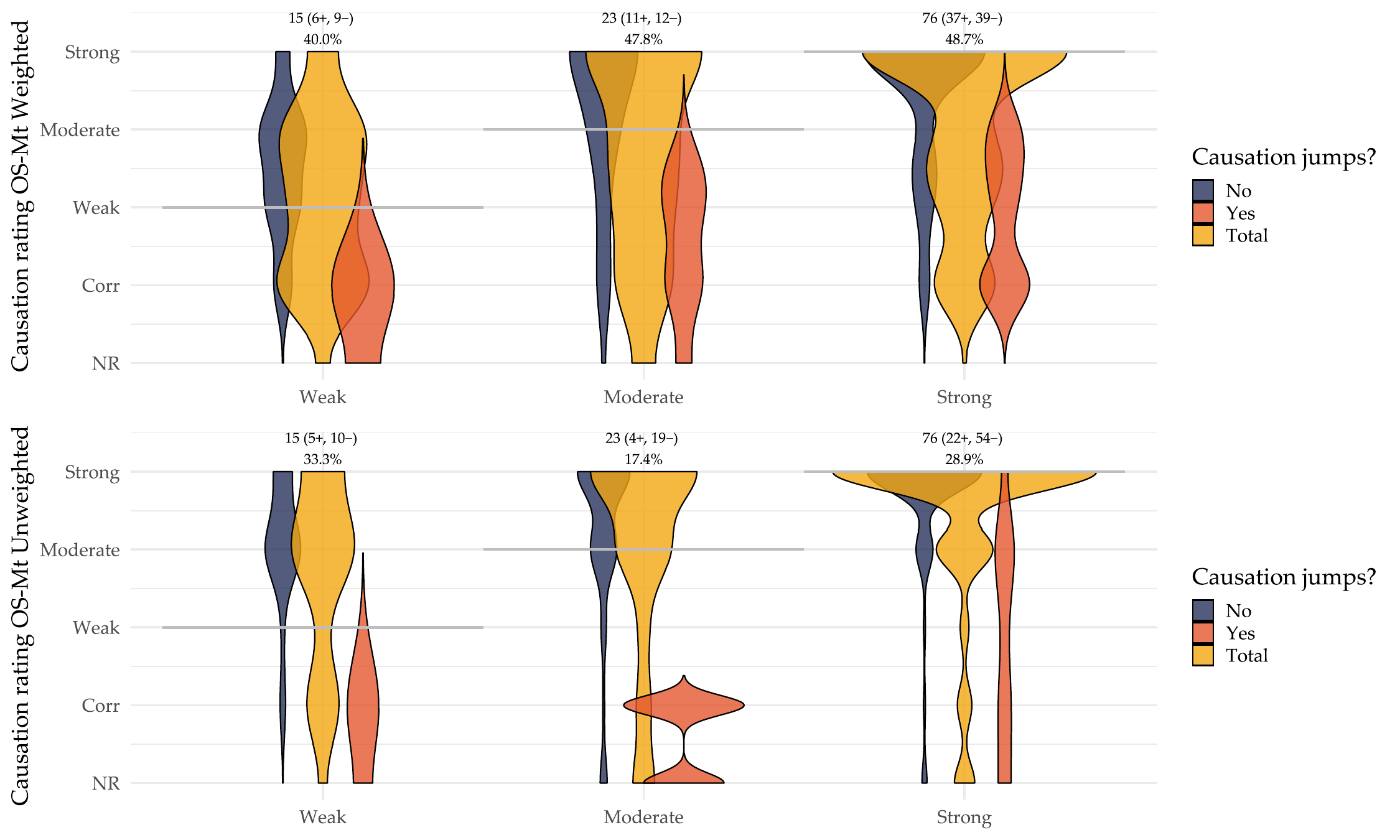}
    \begin{tablenotes}
        \small \textbf{Top panel}: Alignment of causation strength between guideline statements and original study main text conclusive sentences, weighted by content alignment ratings; \textbf{Bottom panel}: Alignment of causation strength between guideline statements and original study main text conclusive sentences, unweighted, original ratings. \\Abbr: NR, no relationship; Corr, correlational; OS-Mt, original study main text conclusive sentence; ``$-$'', causal language jump not present; ``$+$'', causal language jump present. Grey horizontal lines indicate the assumed lowest causation strength required from original studies for each category of guideline statement causation ratings. Numbers and percentages above the panels indicate the total count of guideline statements in each category, and the percentage of causal language jump. 
    \end{tablenotes}
    \caption{Alignment of causation strength ratings}
    \label{fig:alignment-causation}
\end{figure*}

We checked whether causal language jumps were present by comparing the causation ratings in the original studies and in the guideline statements they support (Figure~\ref{fig:heatmap_ratings},\ref{fig:alignment-causation}). Taking content alignment into consideration, 47.4\% (54/114) of the guideline statements could not be effectively supported by any of the associated references and exhibited jumps; even without accounting for content alignment but only comparing the strength of causation, there were still 27.2\% (31/114) of the statements with a causation rating higher than any of their references. This implied potential non-alignment and causal language jumps when passing evidence from the original studies to the guideline statements. In Table~\ref{tab:eg_causation_jump}, we also illustrate different situations where a causal language jump was identified.

The action recommendations contained in the original study abstracts and main texts were also similarly evaluated (Table~\ref{tab:causationdep-rec-guideline}). Half of the abstracts of original studies and two-thirds of the main texts made action recommendations, while more than 90\% of the recommendations in both abstracts and main texts were effectively supported by the conclusions of that original study or did not need evidence support. This differed from the situation in guideline recommendations, implying that guideline recommendations would have required more evidence to support. 

\subsection{Reporting quality of TTE components}
\begin{figure*}[ht]
    \centering
    \includegraphics[width=0.85\textwidth]{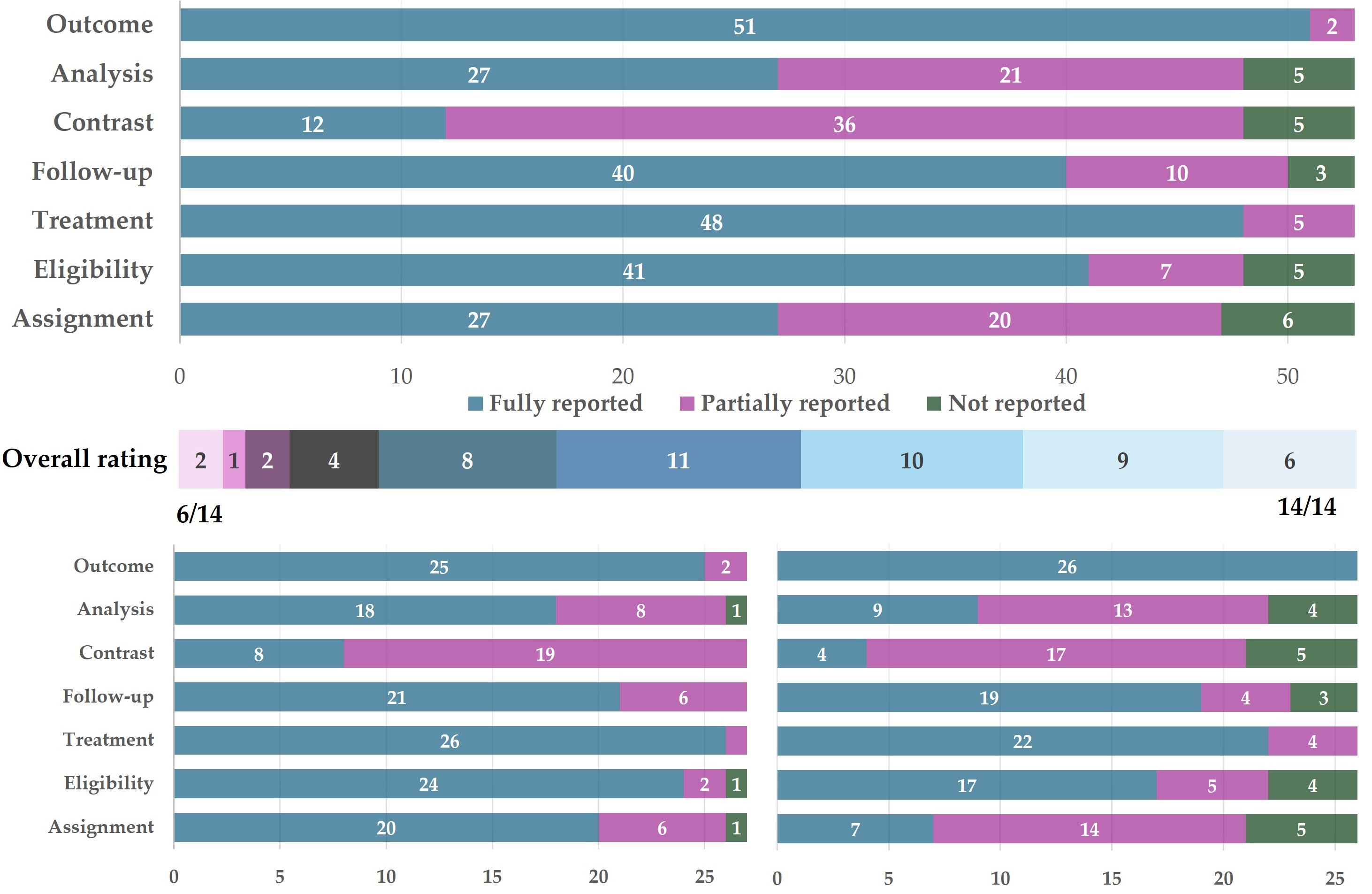}
    \begin{tablenotes}
         \small TTE, target trial emulation. Upper panel illustrates the reporting quality of seven TTE components, and the middle panel gives the overall quality, calculated as the sum of all seven TTE component ratings each ranged from 0 (not reported) to 2 (fully reported). Lower panel gives the details of TTE ratings stratified by study types (lower-left panel for RCT studies, lower-right for non-RCT studies).
    \end{tablenotes}
    \caption{Reporting quality of TTE components}
    \label{fig:TTEcomponent}
\end{figure*}

From all the 97 extracted and further analyzed studies, 53 were primary studies (trials or observational studies) and were eligible for TTE component reporting quality rating. Figure~\ref{fig:TTEcomponent} illustrates the distribution of ratings. Of the seven TTE components, treatment strategy and outcome assessment had the highest reporting quality, whereas causal contrast, analysis plan, and assignment procedure had lower reporting quality. Most studies did not clearly state their causal estimands in the study aims, and implementation details for randomization or covariate adjustment/matching were often not fully given. 

The reporting quality differed by study type (Figure~\ref{fig:TTEcomponent} lower panel). There was a clear gap  between the overall reporting quality scores in RCT (median 12.0, IQR 11.0 to 13.0) and non-RCT studies (median 10.0, IQR 9.00 to 11.0; Wilcoxon test: location shift 2.00 [95\% CI 1.00, 3.00], $W = 556.5$, $P = 0.0002$). Nevertheless, even in clinical trials, the TTE components were not reported at perfect quality.

\subsection{Causal linking words}

We finally characterized the pattern of causal linking word use in both guidelines and original studies. The use of causal linking words in order to express causation varied between guidelines and original studies. In \textbf{File S4} we displayed the most used causal linking words in guideline statements and original studies, in which ``benefit'' was most used in guidelines but ``associated'' took the first place in original studies. ``Associated'' ranked only the fourth by frequencies of all causal linking words in guideline statements. 

\section{Discussion}

In this study, we investigated the use of causal language in selected diabetes guidelines and assessed whether the strength of causation or the dependence to causation was stronger than that in the references they relied on. We characterized the distribution and alignment of causation strength in guideline statements and the original studies they cite, and also assessed the reporting quality of TTE components for eligible primary studies. We revealed that causal language jumps and the non-alignment of expressed strength of causation were common (27.2\% of the statements, or 47.4\% when accounting for contents) between original studies and guideline statements that intended to provide causal evidence to support action recommendations. 

The frequent presence of causal language jumps in guidelines does not necessarily imply that guideline makers made mistakes. For example, despite weaker causation expressed in the cited original studies, a stronger causal conclusion may be warranted when interpreting multiple studies together or in conjunction with other evidence. These jumps must still be made, however, explicitly and on the basis of good causal inference, in the same way as that a single study must make explicit arguments to justify and warrant causal conclusions. Causal language jumps are problematic when they are not warranted (over-interpreting results) or rely on nontransparent reasoning. Though, in this manuscript, we cannot differentiate between ``safe'' and ``dangerous'' jumps; the prevalence of either type of jumps is worth concerning.

Causal language jumps may be present because the strength of causation is hard to determine or could easily be misinterpreted. Controversies about the semantics of different causal/associational words largely exist~\cite{han_causal_2022}: there may even be multiple valid interpretations of a causal sentence. Even for the same causal estimate, two researchers may still choose different causal linking words that agree with their perceptions the most to phrase a statement. While experienced guideline makers may already notice this, the semantic vagueness of causal expressions may still hinder them from getting accurate information. In our study, despite having clear criteria (\textbf{File S2}) and raters specialized in causal inference, divergence between two or more raters was still common. In particular, the strength of causation was harder to judge compared to the presence or absence of causation, as a reflection of subtleties rather than huge differences of the interpretation of a same sentence. 

Another consideration is that original studies with causal intentions or aims often avoided using causal words when concluding. It is common practice in epidemiology to prefer associational language over causal language \cite{kezios_is_2021}, putting the onus on guideline makers to decide whether a causal language jump is warranted. The word ``associated'' particularly exhibits this problem, as it was even ranked the first across all possible causal linking words in original study sentences (\textbf{File S4}), same as that in a previous study~\cite{haber_causal_2022}. That researchers conclude their results as an ``association'' even in a trial, means that they are less likely to analyze potential biases, and quantify post-randomization issues~\cite{hernan_c-word_2018}. However, guideline makers still tended to consider the evidence carried by these studies as stronger causal or as more confident ones, and some jumps were thus made.

Moreover, original studies do not necessarily need to have clear implications for clinical practice, while guidelines are required to do so. Therefore, it may be more necessary for guideline recommendations to make causal language jumps or integrate other domain-specific knowledge to provide definitive conclusions that can be directly used in practice.
As described in \ref{tab:causationdep-rec-guideline}, the causation-dependence of action recommendations was generally stronger than (any of) the corresponding supporting statements. This result was similar to that found in Haber et al. (2022)\cite{haber_causal_2022}. Action recommendations in guidelines were often made not only based on original studies cited in the current single guideline, but also on previous experience, expertise, and other subject-matter knowledge gained from practice and education. These deviations from the original studies may affect the causation-dependence of recommendations and further induce causal language jumps we observed. Incorporating this causal knowledge outside the range of cited original studies is unavoidable and vital, and should be allowed; however, it would be better that scholars with expertise in causal inference and familiar with the disease-specific knowledge participate in this process and help on a safer ``landing'' after this jump.

There are also cases where some studies were indeed incorrectly interpreted or read, and the guideline statements based on these studies thus used incorrect information, including causal information. This happened not only for expressed- and perceived causation, but also for the alignment of contents, for example a specific weight loss goal of 5--7\% was given in guidelines~\cite{committee_5_2024} while the original studies stated weight loss was effective to reduce diabetes incidence but did not give an optimal goal of weight loss~\cite{hamman_effect_2006}. The proportion of causal language jumps doubled after accounting for content alignment compared to the raw causation ratings, implying this could happen more often than supposed.

Lastly, the suboptimal reporting quality of TTE components adds to the ambiguity of the causal information. Evaluation of the quality of the original studies will be harder if some of the TTE components were not reported or insufficiently reported, and this could also potentially lead to causal language jumps on the ``causation flows'' when guideline makers do not read the evidence as intended. Not all the included primary studies in our study reported their TTE components in detail. Most, for example, did not explicitly state their estimands or causal contrasts. This was unexpected given the importance of having a clearly expressed causal question and/or estimand, and would only make the causal claims in the original studies or in guidelines harder to be evaluated. The distribution of TTE component ratings in this study was basically comparable to that reported in Smit et al. (2023) \cite{smit_causal_2023}, with the exceptions that the eligibility criteria were reported at a worse level, and the treatment strategies were better reported. These shifts may come from the clinical trials in this study, while Smit et al. \cite{smit_causal_2023} analyzed only observational studies dedicated to causal inference.

We close our discussion by taking it one step further beyond only guidelines, semantics, or causal \textit{``language''} jump. Actually, each time when causal conclusion is made based on non-causal evidence, a certain ``jump'' has to be made. Instead of a causal \textit{language} jump, we may call this a ``causal jump'' without the word ``language''. In a full causation flow starting at observed data which by nature contain no causation, a causal jump will thus always be required somewhere on the path if the endpoint consists of causal statements. In order to make this jump valid, one must provide justified causal reasoning, and, if necessary, plausible causal assumption sets. Claiming a causal conclusion based on observed data from either an RCT or a cohort also exhibits a jump and requires such reasoning and assumptions. Back to guidelines that claim causation, they should either reference original studies which express causation or provide their reasoning why non-causal research they cite is able to support causal conclusions. 

\subsection*{Limitations}

This study had some limitations. First, we only chose to assess non-pharmacological statements and recommendations in (type 2) diabetes management. Due to our domain-specific knowledge sets and related working experience, we chose to restrict our study to this setting. Second, not all the original studies were fully assessed for TTE reporting due to their large number. Although the assessed studies were chosen randomly, there could indeed be studies of better reporting quality in unsampled references. Third, we assumed that different original studies used to support one guideline statement were independent, and therefore used the study with the strongest causal claim to assess causal jumps. It is possible that several studies each with weaker but related evidence could be synthesize into stronger evidence that suffices to support a stronger causal statement. However, it is not yet possible to look into this ``joint support'' problem at only the semantic level, and the assessment of contents and domain-specific knowledge is beyond the scope of this article.

\section{Conclusion}

Our study revealed that non-alignment between strength of causation and causal language jumps in these causation flows were common in clinical practice guidelines about diabetes management. One third of the guideline statements were causal, and more than one fourth of them made causal language jumps in terms of causation strength, compared to the original studies. The observed jumps can partially be attributed to suboptimal status of causal and associational language use in clinical studies and epidemiology. While causal language jumps are sometimes inevitable to make, good and responsible practice of causal inference can be helpful for a valid and justified jump.

\section{Competing interests}
No competing interest is declared. Dr. J.A. Labrecque is supported by a NWO/ZonMW Veni grant (09150162010213). C.W. is supported by a Chinese Scholarship Council (CSC) scholarship (202307720069). K.W. received no external funding sources. 

\section{Author contributions statement}

The author contributions for this article are: K.W.: Conceptualization, Methodology, Software, Formal analysis, Investigation, Writing - Original Draft, Visualization; C.W.: Validation, Investigation, Writing - Review \& Editing; J.A.L.: Conceptualization, Methodology, Validation, Investigation, Resources, Writing - Review \& Editing, Supervision, Project administration.

\section{Acknowledgments}
We thank prof. dr. Miranda T. Schram from Maastricht UMC+ and Erasmus MC Rotterdam for her suggestions on clinical guideline selection, her expertise in the field of diabetes, and her comments on the diabetes-related contents of this manuscript; we also thank dr. Frank J. Wolters from Erasmus MC Rotterdam for his precious opinions and feedback during the development of this project. 
This work is supported and evaluated by Erasmus MC Graduate School and supporting staff from Erasmus University Rotterdam; we thank them for their support and help. 

\section{Data availability statement}
The pre-registered protocol of this study is available at Open Science Framework (OSF): \url{https://osf.io/25847/resources}. The data that support the findings of this study are available in supplementary files and also upon request from the author (K.W.) if necessary. 

\bibliographystyle{unsrt}
\bibliography{reference}

\end{document}